# Agile Minds, Innovative Solutions, and Industry-Academia Collaboration: Lean R&D Meets Problem-Based Learning in Software Engineering Education

Lucas Romao
Pontifical Catholic University of Rio de Janeiro (PUC-Rio)
Rio de Janeiro, Rio de Janeiro, Brazil
lromao@inf.puc-rio.br

Marcos Kalinowski
Pontifical Catholic University of Rio de Janeiro (PUC-Rio)
Rio de Janeiro, Rio de Janeiro, Brazil
kalinowski@inf.puc-rio.br

Clarissa Barbosa
Pontifical Catholic University of Rio de Janeiro (PUC-Rio)
Rio de Janeiro, Rio de Janeiro, Brazil
cbarbosa@inf.puc-rio.br

Allysson Allex Araújo
Federal University of Cariri (UFCA)
Juazeiro do Norte, Ceará, Brazil
allysson.araujo@ufca.edu.br

Simone D. J. Barbosa
Pontifical Catholic University of Rio de Janeiro (PUC-Rio)
Rio de Janeiro, Rio de Janeiro, Brazil
simone@inf.puc-rio.br

Helio Lopes
Pontifical Catholic University of Rio de Janeiro (PUC-Rio)
Rio de Janeiro, Rio de Janeiro, Brazil
lopes@inf.puc-rio.br

## ABSTRACT

[Context] Software Engineering (SE) education constantly seeks to bridge the gap between academic knowledge and industry demands, with active learning methods like Problem-Based Learning (PBL) gaining prominence. Despite these efforts, recent graduates struggle to align skills with industry needs. Recognizing the relevance of Industry–Academia Collaboration (IAC), Lean R&D has emerged as a successful agile-based research and development approach, emphasizing business and software development synergy. [Goal] This paper aims to extend Lean R&D with PBL principles, evaluating its application in an educational program designed by *ExACTa PUC-Rio* for *Americanas S.A.*, a large Brazilian retail company. [Method] The educational program engaged 40 part-time students receiving lectures and mentoring while working on real problems, coordinators and mentors, and company stakeholders in industry projects. Empirical evaluation, through a case study approach, utilized structured questionnaires based on the Technology Acceptance Model (TAM). [Results] Stakeholders were satisfied with Lean R&D PBL for problem-solving. Students reported increased knowledge proficiency and perceived working on real problems as contributing the most to their learning. [Conclusion] This research contributes to academia by sharing Lean R&D PBL as an educational IAC approach. For industry, we discuss the implementation of this proposal in an IAC program that promotes workforce skill development and innovative solutions.

## KEYWORDS

Problem-Based Learning, Lean R&D, Industry-Academia Collaboration, Innovation, Software Engineering Education

## 1 INTRODUCTION

Software Engineering (SE) education strives to equip students with technical and soft skills [14]. For this reason, SE educators have extensively explored active learning methodologies in pursuit of this objective [19, 27]. In this sense, it has been recognized that active learning methods such as Problem-Based Learning (PBL) can aid software engineers' learning [2, 24]. The underlying principle of PBL posits that the impetus for learning should emanate from the problem the learner endeavors to resolve [5].

However, notwithstanding the progress made through active learning methods, many recent SE graduates still need help with challenges at the outset of their professional journeys, stemming from a misalignment between the skills acquired during their university education and the industry demands [17, 20]. Recognizing this gap, several studies have been undertaken to harmonize SE education with industry needs [1, 13, 18]. Consequently, promoting Industry-Academia Collaboration (IAC) in SE has emerged as a compelling avenue for addressing this disjuncture [4, 7].

To strengthen IAC, Kalinowski et al. [16] have recently proposed an agile Research and Development (R&D) approach for digital transformation, coined as Lean R&D. This proposal was engineered to deliver fast-paced research-intensive Minimum Viable Products (MVP) by leveraging agile, Lean, and continuous SE principles, thereby establishing a robust synergy between business and software development. Notably, the initial evaluation of Lean R&D was confined to the context of R&D projects [15, 29]. However, given its conceptual foundation in IAC, one can contend with the opportunity to adapt and explore Lean R&D within an educational program. This proposition holds particular relevance in SE education owing to the opportunity of exploring a structured approach to harnessing IAC, while simultaneously exposing students to a real-world industrial environment through innovation projects.

Motivated by the aforementioned rationale, this study aims to extend and evaluate the applicability of Lean R&D within an educational program devised by *ExACTa PUC-Rio* for a large-scale Brazilian retail company called *Americanas S.A.*. This Lean R&D extension involved the integration of PBL principles, thereby giving rise to a modified version termed **Lean R&D PBL**. This adaptation encompassed a range of pedagogical practices, including lectures on knowledge tracks in line with typical digital transformation needs, mentoring sessions, and involving students in real-world problem-solving by developing MVPs.

In summary, the educational program following the Lean R&D PBL spanned 12 months, from August 2022 to July 2023. The company's problem suggestions centered on innovations in store management, logistic operations optimization, the application of machine learning for product classification, and e-commerce ad spaces. In collaboration, the program involved 40 undergraduate, master, and PhD students, working part-time (20 hours per week) in four teams. These teams underwent 180 hours of lectures in four digital transformation knowledge tracks and received an additional 180 hours of professional mentoring to address potential challenges.

The Lean R&D PBL assessment took the form of an evaluative case study [26]. The evaluation focused on the perspectives of key stakeholders (students, managers, and sponsors) involved in the educational program. Therefore, we designed a structured questionnaire. The questionnaire collected students' data about the overall Lean R&D PBL acceptance based on the Technology Acceptance Model (TAM) [8], perceived contributions of the pedagogical practices, and knowledge proficiency and company's business understanding before and after the program using the Competencies Proficiency Scale (CPS) [21]. Concerning the managers and sponsors, we also collected their overall acceptance based on TAM and asked them how well the industrial problems were solved.

Our findings indicate that company stakeholders were satisfied with the MVPs and with Lean R&D PBL's problem-solving capabilities. Students reported increased knowledge proficiency in the different knowledge tracks and business understanding. Furthermore, working on real problems was perceived as contributing the most to their learning. This paper discusses the effective integration of PBL principles into Lean R&D, providing valuable insights into designing an IAC educational program.

The remainder of this paper is organized as follows. Section 2 provides an overview of the background and related work. In Section 3, we describe the case study design. In Section 4, we discuss the results. Then, we analyze the limitations and threats to validity in Section 5. Finally, we provide concluding remarks in Section 6.

## 2 BACKGROUND AND RELATED WORK
### 2.1 Problem-Based Learning

PBL stands as a prominent contribution over the past two decades [9]. In summary, PBL manifests as an educational paradigm anchored in problem-solving, providing students with real-world scenarios pertinent to their academic domain. Its extensive utilization in SE education is attributed to its inherent applicability and the fostering of collaborative skills [10, 11, 22, 28].

According to dos Santos et al. [10], there are five key concerns to pay attention to when planning an educational program with PBL. Firstly, the identification and definition of the **Problem** serve as the foundational step, establishing the project to be undertaken by the students. Secondly, creating an optimal **Learning Environment** is emphasized, encompassing both possible physical and virtual spaces to ensure an enjoyable experience for all participants. The third aspect, **Human Capital**, encompasses the diverse stakeholders involved, including students, the educational team, and eventual external stakeholders. Fourthly, guidelines for PBL are articulated under the umbrella of a **Content**, which includes the knowledge, skills, and competencies required to succeed in this field. Lastly, **Processes** encapsulate the varied development approaches and pedagogical classes integral to the overall PBL structure. Together, these five key concerns form a comprehensive approach to successfully implementing PBL.

In the context of SE education, the extensive exploration of PBL is notable. Silva et al. [28], applying PBL in teaching Unified Modeling Language (UML), juxtaposed its efficacy with alternative active learning methodologies. The findings revealed a student predilection for PBL, attributed to its ability to stimulate critical thinking and problem-solving, ease of acquisition, and concomitant generation of more accurate diagrams. Similarly, Olayinka and Stannett [22] employed PBL to teach agile development and business concepts. They found that students acquired technical competencies and developed skills in Scrum practices.

Lastly, dos Santos et al. [11] conducted a Systematic Mapping Study covering PBL in computing studies. The authors have found that PBL is a teaching approach with an optimistic view of the students' resulting in impactful learning. Moreover, while PBL has garnered substantial attention from SE educators, there exists a noticeable research gap in proposals focused on its application in IAC[12]. Applying PBL in IAC projects could facilitate the immersion of the students in real-world industrial environments and their engagement in problems related to real business needs.

### 2.2 Lean R&D

Lean R&D has the purpose of delivering fast-paced Minimum Viable Products (MVPs) within IAC R&D contexts, establishing a robust synergy between business and software development [15]. As shown in Figure 1, Lean R&D comprises four checkpoints, five phases, and the support of a dedicated research team for technical solution-related activities. Hereafter, we summarize these phases to provide an overview. More details can be found in [16].

Lean R&D starts with a *Lean Inception* [6] workshop to allow stakeholders to jointly outline the vision of an MVP that can be used to test business hypotheses. At the first checkpoint, a steering committee (including the sponsor customer) must approve the MVP outline. If the idea of the MVP (or its expected results) is not promising enough, it fails fast, and a new Lean Inception workshop is conducted, potentially focusing on a different business problem.

In the *Technical Feasibility* phase, the development team, assisted by the research team, starts investigating the technical feasibility of implementing the features identified during the Lean Inception. In parallel, the *Conception* involves a Product Owner (PO) detailing the MVP features identified during the Lean Inception by applying product backlog-building co-creation dynamics and other typical agile requirements elicitation and specification techniques (*e.g.*, user stories with acceptance criteria). In this phase, the MVP's User Experience (UX) and User Interface (UI) are also detailed by a UX/UI designer. The second checkpoint involves the steering committee analyzing the requirements specification, mainly based on the UI prototype and the results of the technical feasibility assessment, to decide whether the MVP should be developed.

The *Agile Development* phase involves the development team, with the support of the research team, implementing the MVP following Scrum events. Customer representatives are expected to participate in the sprint planning and reviews. After the development, the third checkpoint involves the PO presenting the MVP to the steering committee so that they can decide upon its transition into production.

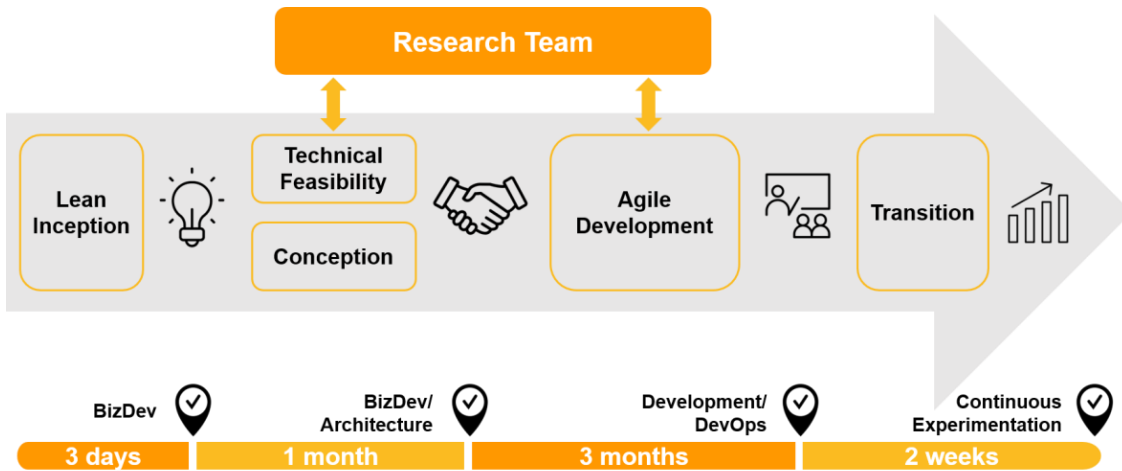

**Figure 1: Overview of the Lean R&D approach [16].**

The *Transition* phase involves the development and infrastructure team preparing the MVP for beta testing in its production environment and assessing the business hypotheses. The last checkpoint concerns analyzing continuous experimentation results to investigate whether the business hypotheses were confirmed and whether it is worth investing in future MVP increments.

## 3 CASE STUDY DESIGN

This investigation concerns an evaluative case study with students, managers, and company stakeholders. We followed the methodological guidelines provided by Runeson et al. [26] for case study research.

### 3.1 Goal and Research Questions

We followed the Goal-Question-Metric (GQM) goal definition template [3] to define our research goal as follows: *Analyze* Lean R&D PBL *for the purpose of* characterization *with respect to* the problem-solving capability, the overall acceptance, the students' perceptions on the pedagogical practices, and students' perceived proficiency in the four knowledge tracks and company's business *from the point of view of* the students and company stakeholders (managers and sponsors) *in the context of* an IAC educational program.

Based on our goal, we established the following Research Questions (RQs):

**RQ1**: How well did the application of Lean R&D PBL solve the proposed problems from the point of view of the company stakeholders?

**RQ2**: What was the acceptance of Lean R&D PBL by the students and company stakeholders?

**RQ3**: How were the different pedagogical practices (lectures, mentoring, and working on real problems) perceived as contributing to students learning?

**RQ4**: How did Lean R&D PBL affect students' perceived proficiency in the four knowledge tracks and the company's business?

### 3.2 Context & Educational Program Design

The context concerns an IAC educational program established between *Americanas S.A.* and *PUC-Rio* that aimed at enhancing the education of undergraduate and graduate students. This program positions students as central figures in the conception and development of innovative technological solutions to address concrete business problems. *Americanas S.A.* is an active enterprise in the Brazilian retail market. We detail the educational program based on the Lean R&D PBL hereafter, aligning it with the five key PBL concerns outlined by dos Santos et al. [10].

*3.2.1 Problem.* As discussed by Dawood and Deriche [9], selecting a problem constitutes a foundational element in the PBL. Within the framework of our educational program, an inherent advantage derived from IAC manifests in identifying problems rooted in our industrial partner's genuine needs. Hence, the program aimed to provide a transformative educational experience immersed in practical industrial applications. We identified four major problems that spanned various domains within our industrial partner's operations. First, enhancing the management of physical stores and the complexities faced by district managers overseeing multiple stores. Second, automatically classifying marketplace products to improve consistency. Third, optimizing ad spaces for products listed on the e-commerce platform. Finally, optimizing delivery routes while maximizing storage capacity.

*3.2.2 Learning Environment.* As shown in Figure 2, the physical learning environment within *ExACTa PUC-Rio* comprised an open workspace featuring workstations for each team and two dedicated meeting rooms. Each student was equipped with a work setup, including a notebook, external monitor, headset, and wireless keyboard and mouse. All lectures, meetings, and showcases occurred within this physical learning environment. The technological infrastructure involved Discord as a communication platform to foster seamless interactions between teams and the company, Jira for agile project management, and GitLab for source code management. A Virtual Private Network (VPN) was also established to enable sensitive data security.

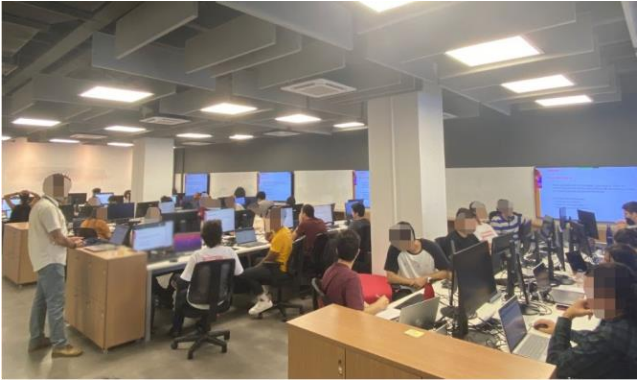

**Figure 2: Physical learning environment at *ExACTa PUC-Rio* during an agile project management lecture.**

*3.2.3 Human Capital.* The human capital of the educational program was multifaceted. It encompassed a pedagogical team from *PUC-Rio*, consisting of coordinators and mentors who contributed to shaping the educational program and provided guidance and support. Furthermore, the program involved managers and sponsors from *Americanas S.A.*, who provided business guidance. Finally, forty students from *PUC-Rio* were selected to participate in the program. We provide further details hereafter.

The pedagogical team comprised the three program coordinators and four mentors from the *ExACTa PUC-Rio* initiative, all holding Ph.D. degrees in distinguished fields such as SE, Data Science, and Human-Computer Interaction. The coordinators took care of the overall program management, overseeing the dedicated student squads within the educational program and facilitating regular meetings with the company. The mentors provided daily guidance to the squads and also provided lectures and mentoring sessions (being complemented by other guest lecturers on demand).

Managers from *Americanas S.A.* represented the company's business problem-related interests during Sprint ceremonies. The sponsors represented the company's overarching strategic interests. Additionally, four representatives of the company's Digital Innovation team helped to coordinate the program and bridge communication.

Concerning the students, the group was formed by 32 undergraduate students, four MSc students, and four PhD students. The majority of them (28) had no experience before enrolling in the program, making it their first industry development experience. The most experienced ones were two students with ten years of experience. Besides them, ten students had 3 ($\pm$1) years of previous experience. These students came from different courses: 12 from Computer Engineering, 11 from Computer Science, seven from Design, two each from Chemical Engineering, Physics, and Production Engineering, and one each from Mechanical Engineering, Neuroscience, Psychology, and Economy. They were grouped into four squads of ten students, containing one PhD student, one MSc student, and eight undergraduate students. Scrum roles were assigned as follows: one student was a Scrum Master, one was a Product Owner, one was a Designer, and the remaining students were developers and data scientists.

*3.2.4 Content.* The educational program was organized into four knowledge tracks, each featuring specialized modules: 1) *Agile Management for Digital Transformation*, comprising Agile Product Management, Agile Project Management, and Ideation Dynamics for Digital Transformation disciplines; 2) *Data Science*, including Data Engineering, Machine Learning, and Advanced Machine Learning disciplines; 3) *Full Stack Development*, covering DevOps, Back-end and Front-end Development disciplines; and 4) *UX/UI Design*, encompassing Ideation Dynamics for Digital Transformation; UX/UI Design, and Data Visualization and Storytelling disciplines. Additionally, an extra module covered the Brazilian General Personal Data Protection Law and Data Governance (LGPD). Each discipline had 36 hours, split into 18 hours for lectures and 18 hours for mentoring. From August 2022 to July 2023, the program comprised two six-month MVP development cycles.

Each student should be enrolled in at least one track. Regarding the distribution of students enrolled in each knowledge track, most students (71%) enrolled in the Data Science track. The Full Stack Development track had an enrollment of 68% of the students. Lastly, the Agile Management for Digital Transformation and UX/UI Design tracks had the lowest enrollment rates, 47% and 34%, respectively.

*3.2.5 Process.* The governance framework for the educational program comprised four essential components, as shown in Figure 3. Following Lean R&D, *Sprint Planning & Review* ceremonies were used for agile project management, and *Checkpoints* were positioned after each phase to validate the accrued business value and to foster a proactive "fail-fast" approach. Additionally, to manage each individual project and the overall program, *Managerial Project Meetings* and *Program Coordination Meetings* were conducted. The *Managerial Project Meetings* were conducted monthly with each project's respective business manager. The *Program Coordination Meetings* were conducted weekly with the company's Digital Innovation team, emphasizing program management and quick decision-making. Moreover, the program coordinators ran a monthly questionnaire to capture students' opinions about what could be improved in the coming months to make the experience more pleasant.

With respect to the process enactment, given the educational focus, we adapted Lean R&D to include lectures and mentoring in the project dynamics while working on the problems. Each week the students had four hours of lectures and four hours of on-the-job mentoring (32 hours of lectures and mentoring per month). The first weeks of the program were slightly more lecture-intensive to properly prepare the students to start working on their tasks.

### 3.3 Data Collection

Data collection covered perceptions from the stakeholders (students, company managers, and sponsors) involved in the educational program. We collected students' data about the overall acceptance, perceptions on the pedagogical practices and on their proficiency in the knowledge tracks and the company's business understanding. Therefore, we designed a structured questionnaire incorporating three parts: 1) an adaptation of the Technology Acceptance Model (TAM) [8], 2) an evaluation of the pedagogical practices' effectiveness for learning, and 3) a self-evaluation before and after the program following the Competencies Proficiency Scale (CPS) [21].

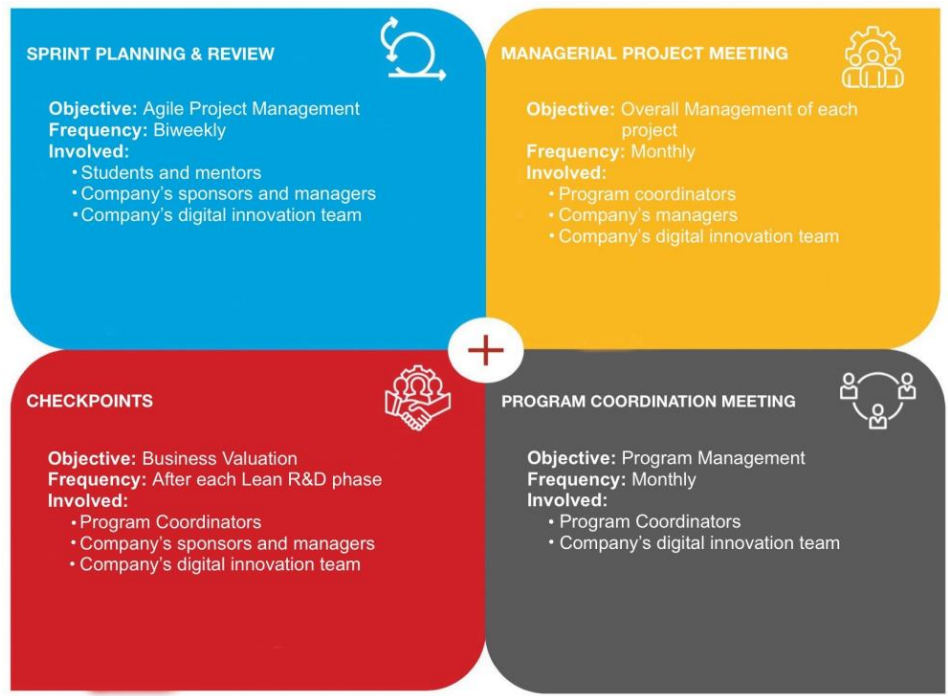

Figure 3: Educational program governance.

Regarding the overall acceptance, TAM is a widely recognized model for evaluating acceptance of new technologies [30]. Hence, we used the TAM statements on the *Perceived Usefulness*, *Perceived Ease of Use*, and *Intent to Use* to be assessed using a five-point Likert scale, ranging from complete disagreement to complete agreement.

For the pedagogical practices, our educational program comprised the following three: addressing real-world problems, lectures, and mentoring sessions. Our questionnaire included a dedicated section with three closed-ended questions using a similar five-point Likert scale to assess the students' agreement on each practice contributing to their learning. We also collected qualitative feedback using open-ended questions.

Finally, we used the CPS scale to determine the students' perceived proficiency in the knowledge tracks and in the company's business understanding before and after the program. The progression of skills within this scale can be delineated as follows: at the *Apprentice level*, individuals possess an awareness of fundamental concepts but lack practical experience. Moving to the *Beginner stage*, one has limited practical experience, typically gained through classroom projects, and may require assistance. On the *Intermediate level*, individuals can apply their knowledge in practice, occasionally seeking assistance. Progressing further to the *Advanced level*, individuals can adeptly apply their skills in practice without needing assistance and are proficient enough to assist others. Finally, at the *Expert level*, individuals consistently demonstrate excellence in practical application across various projects and organizations, establishing themselves as a recognized reference in the field. Again, we also collected qualitative feedback using open-ended questions.

For company managers and sponsors, we assessed only their overall acceptance, given that they were not the target audience for the educational questions. It is noteworthy that we had signed a collaboration contract with a non-disclosure agreement and explicit mentions of shared interest in research. Therefore, we only collected data on the practical application and benefits of the educational program. We also informed the participants that answering the questionnaires was optional and that their data would be anonymized.

We received a total of 42 responses out of 46 invitations: 38 from students, two from company managers, and two from company sponsors. For full access to the questionnaires and the anonymized raw data, please refer to the supplementary materials, which can be found in our open science repository [25].

### 3.4 Data Analysis

We conducted quantitative analyses employing descriptive statistics for closed-ended questions. We analyzed the outcomes of our questionnaire, covering the TAM section, students' perceptions of different pedagogical practices, and self-evaluation of knowledge proficiency and company business. The collected data was visually presented using stacked bar graphs, enhancing result interpretation. For the quantitative examination of perceived knowledge proficiency, we compared responses before and after the program, excluding those marked as "not applicable". The valid results, categorized into the CPS scale Beginner, Apprentice, Intermediate, Advanced, and Expert levels, were also depicted on a stacked bar graph.

Given the overseeable number of open-text responses, we did not apply more sophisticated qualitative analysis methods. However, as recommended for case study research [23, 26], we complement our analyses by using quotations from the open-text questions to illustrate the findings.

## 4 CASE STUDY RESULTS AND DISCUSSION

### 4.1 RQ1: Problem solving capability

The Lean R&D PBL helped us to orchestrate the learning process, encompassing all stages of a software development cycle. The approach provided students with an immersive real-world experience. Furthermore, the application of Lean R&D PBL enabled us to deliver five MVPs, adhering to the predefined timelines, with deliveries in January and July 2023. In the end, two MVPs were deployed at the industry partner.

The squads developed an "Intelligent Management App for Stores" mobile app to address the first problem concerning enhancing the management of physical stores. The project reached a significant milestone by successfully delivering the first MVP in January 2023, providing substantial support to store managers in their daily tasks. This deployment prompted the generation of valuable ideas for further enhancements, pursued throughout the first semester of 2023. These efforts culminated in conceptualizing and developing a second MVP related to the same problem, a web application designed to assist district managers in overseeing their stores. By July 2023, both MVPs had been effectively delivered to the development teams of the Industrial Partner. This project exemplifies Lean R&D PBL enabling the students to translate conceptualizations into tangible solutions that address real-world problems.

Additional MVPs were dedicated to addressing the problems of automatic classification of seller products in the company's marketplace, optimizing ad spaces for products listed on the e-commerce platform, and optimizing delivery routes while maximizing storage capacity. Despite the successful delivery of these MVPs, including solutions involving machine learning and sophisticated optimization techniques, meeting the requirements and goals established during the Lean Inception, they were archived. This decision was motivated by shifts in *Americanas S.A.*'s strategic focus at the beginning of 2023, prioritizing physical stores.

Regarding feedback, one sponsor expressed satisfaction with the Lean R&D PBL approach, emphasizing its suitability for outlining MVPs and developing products. She mentioned that "Lean R&D was very appropriate for the projects, not just because it made them agile, but because it placed all participants as protagonists in understanding the problem, ideating the solution, and building the MVP, fundamental steps to favor innovative solutions". Another sponsor highlighted that "the Lean R&D approach helped to structure the projects, from conception to development, providing a methodology for integrating the business areas and the research and development teams". This acknowledgment underlines the effectiveness of Lean R&D PBL in facilitating collaboration and ensuring a comprehensive methodology throughout the project lifecycle.

The company's satisfaction with the overall program and its problem-solving capability is further evident through a public communication post[1] shared by the company on its LinkedIn profile to more than one million followers. The company expressed, "The project aimed to allow students to find solutions to real-world problems and enrich their training, combining it with graduate-level research and direct exposure to the company's specific challenges. The projects were carried out with great efficiency and dedication by the students and the professors from *PUC-Rio* in partnership with the business and digital innovation teams of *Americanas S.A.*".

Feedback from the company's stakeholders indicates that the Lean R&D PBL enabled effectively addressing the business problems, producing MVPs that are now in use at the company's stores. The collaboration with our partner enabled the identification of genuine problems, an essential component for the success of any PBL program [11, 13]. Additionally, the implementation of our educational program reinforced key factors for successful Industry-Academia Collaboration (IAC), as identified by Wohlin et al. [32], including effective collaboration, alignment with business strategy, commitment to contributing to the industry, and the availability of a highly qualified research environment within the university.

**RQ1 Answer:** Based on our analysis, it is possible to conclude that the application of Lean R&D PBL was suitable and able to solve the proposed problems from the point of view of the company's stakeholders.

### 4.2 RQ2: Overall acceptance

As discussed in Section 3, participants were surveyed to measure their level of agreement regarding Perceived Usefulness, Perceived Ease of Use, and behavioral Intent to Use (*i.e.*, constructs from TAM) after adopting Lean R&D PBL. The obtained results from students are illustrated in Figure 4, while Figure 5 presents the outcomes from the company's managers and sponsors.

Regarding Figure 4, most responses indicated agreement with the Lean R&D PBL approach based on the TAM, reflecting an overall positive perception for all constructs. The two questions that were most disagreed with refer to ease of learning (Q5) and ease of applying (Q6), both related to the perceived Ease of Use construct. This issue is understandable, given that many participants were engaging in their first real-world software development cycle from conception to deployment. Still, the disagreement reached only 14% for Q5 and 11% for Q6, while 71% agreed for both questions.

Regarding open feedback about the ease of applying the Lean R&D PBL, one student discussed that the team faced bureaucratic delays during the Technical Feasibility and Conception phases, leading to insecurity and demotivating the team. This issue, though causing discomfort, was attributed not to the Lean R&D PBL easiness to apply but rather to internal security agreement delays in sharing sensitive information. Once these agreements were in place, the Technical Feasibility and Conception phases proceeded as expected.

Concerning the company's sponsors and managers, Figure 5 revealed that the managers and sponsors also largely agreed with the TAM constructs. However, one of the two managers positioned himself neutrally regarding the ease of use (Q7), and one of the two sponsors reported a neutral position on improved productivity (Q2) caused by the Lean R&D PBL.

---

[1] https://www.linkedin.com/posts/americanas-sa_americanas-futuro-lab-puc-rio-activity-7100866017156841472-Nvmo?utm_source=share&utm_medium=member_desktop

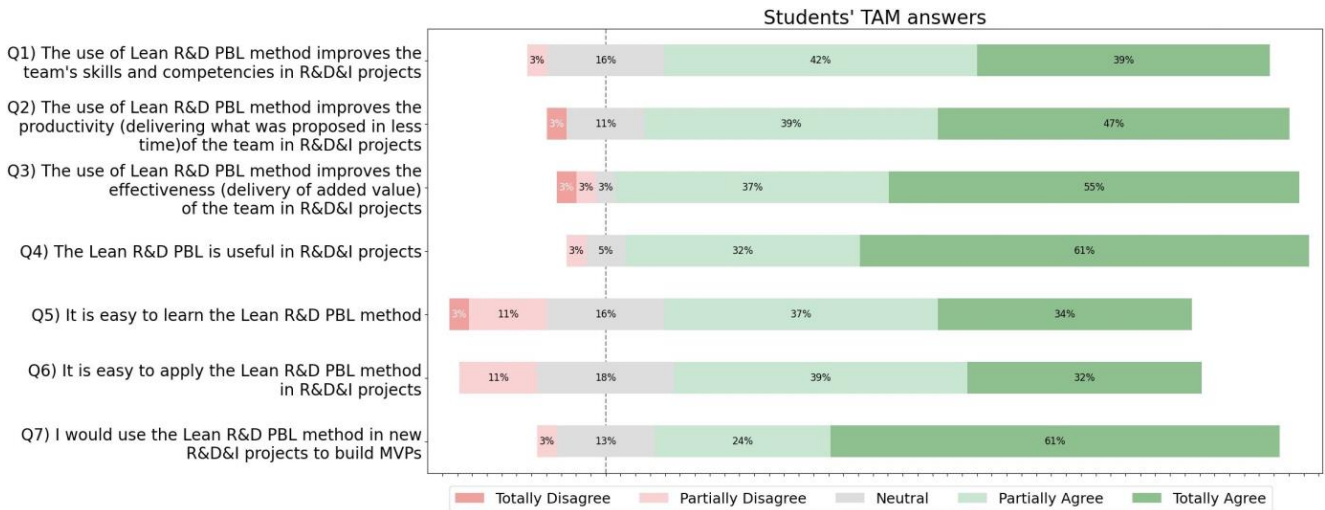

Figure 4: Students responses to the TAM questions.

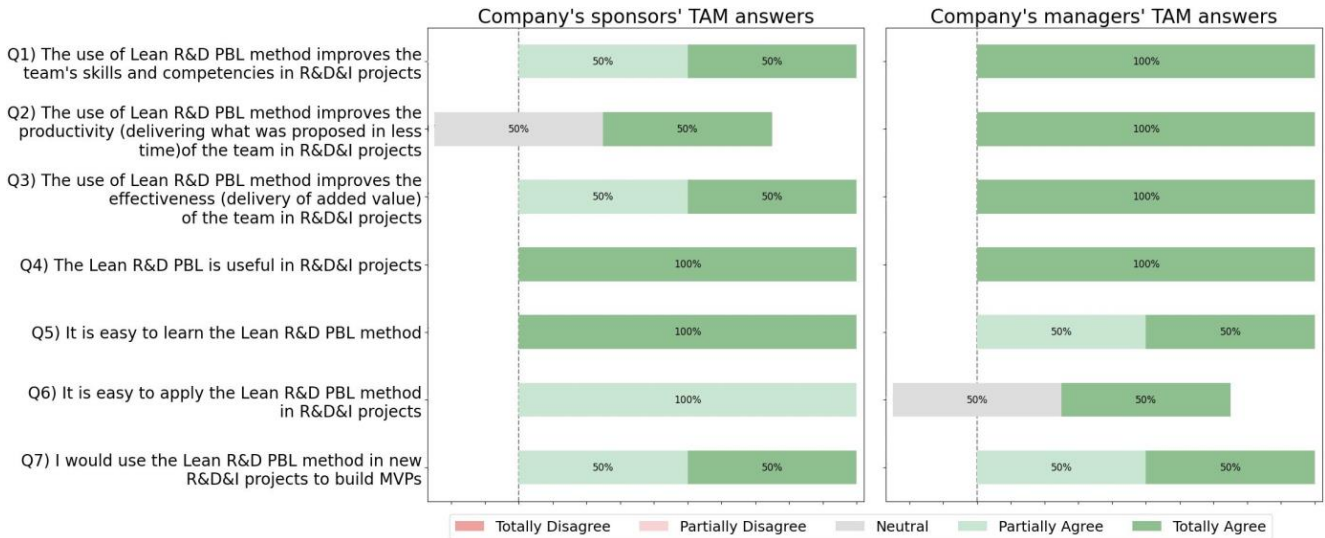

Figure 5: Sponsors and managers responses to the TAM questions.

As we can see, even with minor disagreements, Lean R&D PBL received significant overall acceptance from our industrial partner and students. We found that the approach was able to deliver short-term results with benefits to the industrial partner.

The Lean R&D PBL engaged students in activities across different phases of software development. These activities contribute to developing skills identified as highly valued in the industry [13]. Furthermore, the Lean R&D PBL facilitated regular interactions between students and company stakeholders and ensured the research team's efforts aligned with the industrial team's needs. These aspects are highlighted as key for fostering alignment between industry and academia [7].

**RQ2 Answer:** After analyzing the data, we can conclude that Lean R&D PBL was well accepted by both students and stakeholders of the industrial partner.

## 4.3 RQ3: Perception about the pedagogical practices

As clarified in Section 3, each knowledge track encompassed three pedagogical practices. Firstly, traditional lectures involve lecturers sharing knowledge while students engage through note-taking and questions. Secondly, mentoring sessions allowed students to seek guidance, fostering active learning in teamwork. Lastly, real-world problem-solving allowed students to put theoretical concepts into practice.

Figure 6 illustrates students' opinions on how each pedagogical practice contributed to their learning. Our analysis revealed that seven students (21%) expressed some dissatisfaction with their experience in the lectures. For example, one student mentioned that "the tracks have a very good intention; however, the lectures are superficial to suit everyone, leaving them monotonous and even unnecessary for the more experienced." Another feedback provided by one of the students was the following: "Lectures often did not follow our project schedule. Therefore, the content applied in classes did not converge with our demands". These students expressed a desire for more complex content or topics that more directly align with their project requirements.

The initial aim of the classes was to provide a basis for students with varying levels of knowledge, including those completely new to computer science, SE, or UX/UI. As indicated in Figure 6, we observed that a considerable part (47%) of students agreed that lectures contributed to their learning. In summary, we recognize that the lectures may benefit from a reformulation for our next educational program cycle.

Regarding their experience with mentoring, only one student (3%) reported that mentoring did not contribute to their learning. However, this student clarified that his dissatisfaction was more related to the beginning of the program: "At the end of the cycle, it was adjusted, but I think the mentoring ended up being out of step with the challenges; sometimes we had mentoring available for problems that we had already solved, or sometimes it was the opposite". In fact, we continuously improved our mentoring strategies based on students' feedback throughout the program, adjusting them to better suit their projects.

In general, most students (84%) positioned themselves positively regarding the contribution of mentoring to their learning. Students have found mentoring to be useful and conducive to their learning experience. They considered professional guidance fundamental, with one student stating that "Mentoring was fundamental". Another student expressed satisfaction with the opportunity to engage with professionals, mentioning that "More than the classes, being able to discuss with professionals and receive tips was the most valuable aspect of the learning proposals". In particular, students highlighted the positive experience assigned to the mentoring in Agile Management for Digital Transformation. Students, especially the Product Owners and Scrum Masters, shared a common sentiment, as expressed in the following statement made by one of them: "The Agile methodology mentoring session was very useful for the development of the project. It clarified concepts that were not in consensus among the team, such as defining tasks, user stories, frequency of commits, branches, etc". More than lectures alone, professional guidance can support in providing practical instruction and enhancing the overall learning experience.

Finally, we observed that all students agreed that real-world problem-solving contributed to their learning. The open feedback demonstrates the impact of our educational program on the students' learning. For some students, it was an opportunity to apply their theoretical knowledge in practical scenarios, as one expressed: "Before this lab, I could not imagine how I would work as a developer. I was just studying and learning more skills during college, without knowing how I would use them in real life, but this project was very helpful for me to have that experience".

Indeed, one of the main features of the program is to enable students to merge their academic learning with industry needs. This feature was valued by one student: "Real challenges provide an experience that classroom or personal projects lack. Possibly poorly organized folders, client pressure to deliver, sprints. All these are elements that exist in the industry but are not seen in solo projects". Other students also provided further feedback on this contrast. One student remarked, "Working on the challenges was extremely important for my learning, such as how we should deal with customers and how we should organize the project and the team". Another student shared how their project can impact real lives: "Despite being at an internship level, it is practically a real work experience in the market [...] every step and 'well done, team' heard from the customers was a great satisfaction and rewarding to hear. We believe that many jobs will be positively affected by our work as well."

SE education has been emphasizing the importance of active learning methods [9, 11, 28]. The feedback from students regarding the pedagogical practices adopted within the Lean R&D PBL experience indicated a clear preference for scenarios in which they were positioned as active participants rather than lectures, for example. This preference highlights the impact of active engagement on the learning process perception, suggesting that active methods tend to enhance learning.

**RQ3 Answer:** Based on our analysis, we can conclude that the pedagogical practices (lectures, mentoring, and working on real-world problems) contributed to students learning. However, mentoring and working on real-world problems were perceived as more suitable than lectures.

### 4.4 RQ4: Perceived proficiency in the knowledge tracks and company's business

Figure 7 shows the student's self-evaluation proficiency on each knowledge track following the CPS scale, before and after the program. Hereafter we discuss the results for each knowledge track.

*Data Science*: The program's impact on Data Science proficiency was evident through significant shifts in knowledge levels. Initially, a majority (67%) fell into the *Apprentice* or *Beginner* categories, with only a small slice (4%) at the *Advanced* level. After the program, 74% of the participants self-evaluated themselves from *Intermediary* to *Expert*.

*Full Stack Development*: Similar to Data Science, the majority (72%) of students considered themselves as *Apprentice* or *Beginner* before the program. In turn, 66% of participants self-evaluated themselves from *Intermediary* to *Expert* after the program.

*Agile Management for Digital Transformation*: The proficiency in this track reached the highest difference when comparing the results before and after the program. By analyzing the results before the program, we observed that a large majority (94%) of participants were at the *Apprentice* or *Beginner* level. After the program, the shift was considerable, with 76% of students self-evaluating themselves from *Intermediary* to *Expert*.

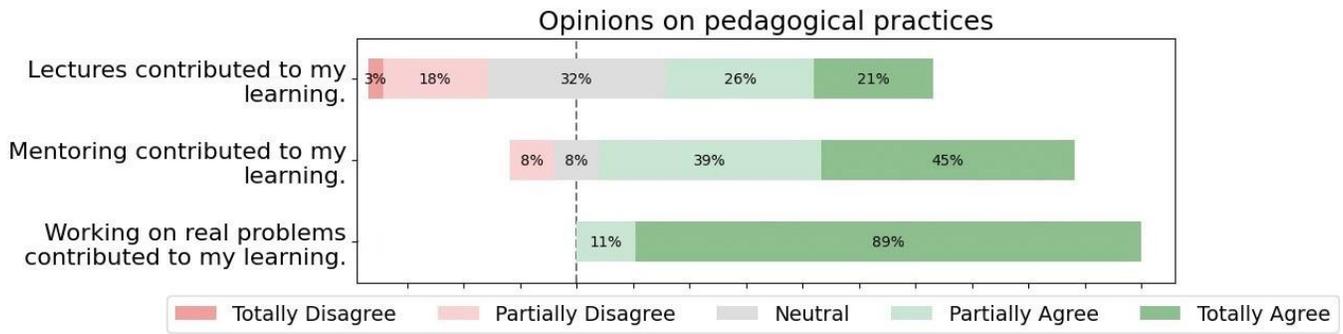

Figure 6: Students' opinions on how each of the pedagogical practices contributed to their learning.

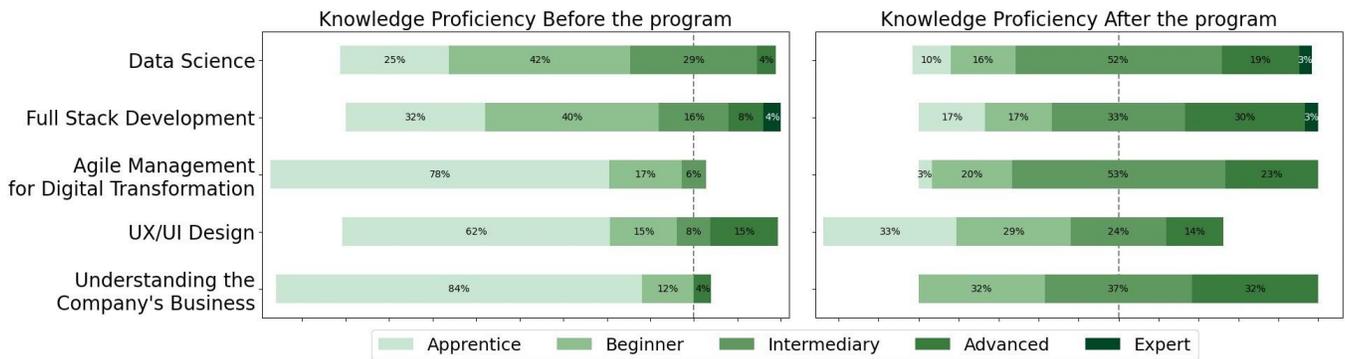

Figure 7: Students' knowledge proficiency before and after the program.

*UX/UI Design*: This track encompassed a smaller number (13) of enrolled students and only four of the respondents took an active part in UX/UI design activities in the projects. Therefore, the proficiency shift before and after the problem was naturally lower. Still, it is possible to see some improvement, given that the number of participants declaring themselves as *Apprentice* approximately decreased by half after the program.

*Understanding the Company's Business*: This perspective also achieved a highly positive proficiency shift when comparing the results before and after the program. We noticed that 96% of the students considered themselves as *Apprentice* or *Beginner* before the program, while 69% of them self-evaluated from *Intermediary* to *Expert* after the program.

Regarding open-text answers, one student expressed, "I notice that I have more fluidity working with data science, without resorting to the internet all the time, and sometimes I helped colleagues during the project." Adding to their perspectives, one student claimed, "I had no idea how things worked properly, but knowing the role of Scrum Master, PO, how the MVP should be done – this was one of the coolest pieces of knowledge in the program, to the point that I became Scrum Master of my squad for over a month". According to Vygotsky [31], incorporating the social context is essential for attaining higher mental functions and fostering learning. In our program, we observed that learning was not solely an individual endeavor; there was a significant exchange of experiences among students. As one student evidenced, "Here and there, I was able to help my colleagues with the work I had done".

Despite some students having prior experience with agile development, they found their experience in the program more enriching than their previous experiences. One student noted, "I had experience in full stack development from my previous internship, but I feel that this project allowed me to develop certain tasks more independently, creating confidence to help others with their tasks". Another student highlighted, "I had previous contact with agile methodology; however, I learned even more in the program. My maturity occurred with the application of the method in practice using real problems of a company as a foundation".

Overall, the students' open-text responses highlighted the perceived value of working on real projects while learning. This issue is also discussed by Garousi et al. [13], where they emphasize that topics should be aligned with real problems in a real-world context to enhance learning outcomes. Indeed, this alignment between industry expectations and what is provided by formal education for internship-level employees is quite challenging [2].

**RQ4 Answer:** Upon analyzing the data, it was found that most students perceived improvements in their knowledge proficiency across all knowledge tracks and their understanding of the company's business.

## 5　THREATS TO VALIDITY

In this section, we discuss threats to the validity of our study according to the categories presented by Runeson et al. [26].

In terms of **construct validity**, potential challenges include the influence of *evaluation apprehension* that can mobilize the students to avoid negative feedback. To mitigate this threat, the questionnaire was applied in an anonymous way. Additionally, *hypothesis guessing* emerges as a potential threat, where participants may have provided overly positive responses anticipating the research being conducted. To minimize this issue, we clarified that the questionnaire would be used to investigate ways to understand and improve the students' experience.

For **internal validity**, issues such as *diffusion or imitation of treatments* come to the forefront. Some students may have sought information about the motivation for some responses from their peers, compromising the integrity of our study. To minimize this threat, we included open questions at the end of each question to complement the quantitative analysis of the results and better understand students' perceptions when answering each question. Furthermore, the presence of *resentful demoralization* introduces the risk of negative evaluations from some participants who may harbor dissatisfaction with some of the coordinators during their experience. To address this, we checked the distribution of the answers of the participants and also analyzed the open-text answers.

Regarding **external validity**, we faced the *interaction of selection and treatment* due to the non-representative sample reached by our study. With respect to the company stakeholders, although we had a small number of company respondents, all of them played important and strategic roles. Concerning the students' sample, even with a high percentage (95%) of students' answers, we recognize the importance of running more cycles of Lean R&D PBL with different students toward generalization.

Finally, with respect to **reliability**, we conducted a single case study. However, as expected for case studies, we carefully describe the context, and our case study fills a literature gap by exploring PBL approaches within an IAC context with a large Brazilian company in the retail market. To improve the reliability of our analyses, we peer-reviewed all our analysis procedures involving an independent researcher and made all data available online for auditing in our open science repository [25].

## 6　CONCLUSION

PBL has been used to improve SE education but often fails to shorten the gap between academic learning and industry needs. IAC emerges as a potential solution, aligning educational objectives with market demands. This paper presents our combined application of Lean R&D and PBL in collaboration with a large Brazilian retail company.

We conducted an evaluative case study focusing on the empirical analysis of four aspects regarding Lean R&D PBL: (1) the problem-solving capability, (2) the overall acceptance, (3) the students' perception of the pedagogical practices, and (4) students' perceived proficiency in the four knowledge tracks and company's business before and after the program. Therefore, we designed a semi-structured questionnaire based on the Technology Acceptance Model (TAM) and the Competencies Proficiency Scale (CPS). Answers were collected from students and industry partners (managers and sponsors).

The findings of our study indicate that the company stakeholders reported satisfaction with the problem-solving capability of the Lean R&D PBL approach. Furthermore, Lean R&D PBL was largely accepted by the company stakeholders and students. Real-world problem-solving was the pedagogical practice perceived as contributing the most to students' learning, followed by mentoring and lectures. We identified a need to restructure the lectures to better align with the industrial problems being addressed by the students. Importantly, students perceived considerable proficiency gains in all four knowledge tracks, alongside an improved understanding of the company's business domain.

This paper contributes to academia by advancing the discussion on innovative educational methodologies in SE through the integration of PBL principles in an IAC using Lean R&D. We also offer practical implications for industry by showcasing the successful application of Lean R&D PBL in a collaborative educational program with a large-scale company. The Lean R&D PBL approach can be applied in IAC contexts, such as capstone projects or internships. The Minimum requirements include active participation from industry partners, a well-defined problem scope, and adequate resources for the overall program coordination, teaching, and mentoring. For future work, we aim to conduct a multiple case study also considering data from other projects at *ExACTa PUC-Rio* involving different industrial partners that are also using Lean R&D in PBL contexts.

## ARTIFACT AVAILABILITY

All the artifacts underpinning this study are openly available via our online open science repository [25].

## ACKNOWLEDGEMENTS

We thank *Americanas S.A.*, including the program sponsors, business managers, and especially the Digital Innovation team, for all the support in establishing a fruitful industry-academia collaboration. We also thank the students, mentors, and lecturers enrolled in the program for their commitment to helping to make this collaborative educational program such a great learning experience for all the involved parties. We thank the Brazilian Council for Scientific and Technological Development (CNPq process #312275/2023-4) for financial support.

## REFERENCES

[1] Nurul Ezza Asyikin Mohamed Almi, Najwa Abdul Rahman, Durkadavi Purusothaman, and Shahida Sulaiman. 2011. Software engineering education: The gap between industry's requirements and graduates' readiness. In *2011 IEEE Symposium on Computers & Informatics*. IEEE, 542–547.
[2] Jocelyn Armarego. 2007. Beyond PBL: preparing graduates for professional practice. In *20th Conference on Software Engineering Education & Training (CSEET'07)*. IEEE, 175–183.
[3] Victor R Basili and H Dieter Rombach. 1988. The TAME project: Towards improvement-oriented software environments. *IEEE Transactions on software engineering* 14, 6 (1988), 758–773.
[4] Kathy Beckman, Neal Coulter, Soheil Khajenoori, and Nancy R Mead. 1997. Collaborations: closing the industry-academia gap. *IEEE software* 14, 6 (1997), 49–57.
[5] David Boud. 1985. *Problem-based learning in education for the professions*. Higher Education Research and Development Society of Australasia.
[6] Paulo Caroli. 2017. *Lean inception*.
[7] Jeffrey C Carver and Rafael Prikladnicki. 2018. Industry–academia collaboration in software engineering. *IEEE Software* 35, 5 (2018), 120–124.
[8] Fred D Davis. 1989. Perceived usefulness, perceived ease of use, and user acceptance of information technology. *MIS quarterly* (1989), 319–340.
[9] Anwar Dawood and Mohamed Deriche. 1999. Riding the wave of new strategies in engineering education. In *ISSPA'99. Proceedings of the Fifth International Symposium on Signal Processing and its Applications (IEEE Cat. No. 99EX359)*, Vol. 2. IEEE, 555–558.

[10] Simone C dos Santos, Felipe Furtado, and Walquiria Lins. 2014. xPBL: A methodology for managing PBL when teaching computing. In *2014 IEEE Frontiers in Education Conference (FIE) Proceedings*. IEEE, 1–8.

[11] Simone C. dos Santos, Priscila B. S. Reis, Jacinto F. S. Reis, and Fabio Tavares. 2021. Two Decades of PBL in Teaching Computing: A Systematic Mapping Study. *IEEE Transactions on Education* 64, 3 (2021), 233–244. https://doi.org/10.1109/TE.2020.3033416

[12] Awdren Fontão, Edson Matsubara, Henrique Mongelli, Marcio Medeiros, Carlos Lourenço, Henrique Martins, Igor Cortez, and Maria Borges. 2023. Hyacinth macaw: a project-based learning program to develop talents in Software Engineering for Artificial Intelligence. In *Proceedings of the XXXVII Brazilian Symposium on Software Engineering*. 312–321.

[13] Vahid Garousi, Gorkem Giray, Eray Tuzun, Cagatay Catal, and Michael Felderer. 2019. Closing the gap between software engineering education and industrial needs. *IEEE software* 37, 2 (2019), 68–77.

[14] Scott Heggen and Cody Myers. 2018. Hiring millennial students as software engineers: a study in developing self-confidence and marketable skills. In *Proceedings of the 2nd International Workshop on Software Engineering Education for Millennials*. 32–39.

[15] Marcos Kalinowski, Solon Tarso Batista, Helio Lopes, Simone Barbosa, Marcus Poggi, Thuener Silva, Hugo Villamizar, Jacques Chueke, Bianca Teixeira, Juliana Alves Pereira, Bruna Ferreira, Rodrigo Lima, Gabriel da Silva Cardoso, Alex Furtado Teixeira, Jorge Alam Warrak, Marinho Fischer, André Kuramoto, Bruno Itagyba, Cristiane Salgado, Carlos Pelizaro, Deborah Lemes, Marcelo Silva da Costa, Marcus Waltemberg, and Odnei Lopes. 2020. Towards lean R&D: an agile research and development approach for digital transformation. In *2020 46th Euromicro Conference on Software Engineering and Advanced Applications (SEAA)*. IEEE, 132–136.

[16] Marcos Kalinowski, Hélio Lopes, Alex Furtado Teixeira, Gabriel da Silva Cardoso, André Kuramoto, Bruno Itagyba, Solon Tarso Batista, Juliana Alves Pereira, Thuener Silva, Jorge Alam Warrak, Marcelo da Costa, Marinho Fischer, Cristiane Salgado, Bianca Teixeira, Jacques Chueke, Bruna Ferreira, Rodrigo Lima, Hugo Villamizar, André Brandão, Simone Barbosa, Marcus Poggi, Carlos Pelizaro, Deborah Lemes, Marcus Waltemberg, Odnei Lopes, and Willer Goulart. 2020. Lean r&d: An agile research and development approach for digital transformation. In *Product-Focused Software Process Improvement: 21st International Conference, PROFES 2020, Turin, Italy, November 25–27, 2020, Proceedings 21*. Springer, 106–124.

[17] Marco Kuhrmann, Joyce Nakatumba-Nabende, Rolf-Helge Pfeiffer, Paolo Tell, Jil Klünder, Tayana Conte, Stephen G MacDonell, and Regina Hebig. 2019. Walking through the method zoo: does higher education really meet software industry demands?. In *2019 IEEE/ACM 41st International Conference on Software Engineering: Software Engineering Education and Training (ICSE-SEET)*. IEEE, 1–11.

[18] Claire Le Goues, Ciera Jaspan, Ipek Ozkaya, Mary Shaw, and Kathryn T Stolee. 2018. Bridging the gap: From research to practical advice. *IEEE Software* 35, 5 (2018), 50–57.

[19] Bruce R Maxim, Adrienne Decker, and Jeffrey J Yackley. 2019. Student Engagement in Active Learning Software Engineering Courses. In *2019 IEEE Frontiers in Education Conference (FIE)*. IEEE, 1–5.

[20] Nancy R Mead. 2015. Industry/university collaboration in software engineering education: refreshing and retuning our strategies. In *2015 IEEE/ACM 37th IEEE International Conference on Software Engineering*, Vol. 2. IEEE, 273–275.

[21] National Institute of Health. 2009. The NIH Proficiency Scale. https://hr.nih.gov/working-nih/competencies/competencies-proficiency-scale

[22] Olakunle Olayinka and Mike Stannett. 2020. Experiencing the Sheffield Team Software Project: A project-based learning approach to teaching Agile. In *2020 IEEE Global Engineering Education Conference (EDUCON)*. 1299–1305. https://doi.org/10.1109/EDUCON45650.2020.9125175

[23] Paul Ralph, Nauman bin Ali, Sebastian Baltes, Domenico Bianculli, Jessica Diaz, Yvonne Dittrich, Neil Ernst, Michael Felderer, Robert Feldt, Antonio Filieri, Breno Bernard Nicolau de França, Carlo Alberto Furia, Greg Gay, Nicolas Gold, Daniel Graziotin, Pinjia He, Rashina Hoda, Natalia Juristo, Barbara Kitchenham, Valentina Lenarduzzi, Jorge Martínez, Jorge Melegati, Daniel Mendez, Tim Menzies, Jefferson Molleri, Dietmar Pfahl, Romain Robbes, Daniel Russo, Nyyti Saarimäki, Federica Sarro, Davide Taibi, Janet Siegmund, Diomidis Spinellis, Miroslaw Staron, Klaas Stol, Margaret-Anne Storey, Davide Taibi, Damian Tamburri, Marco Torchiano, Christoph Treude, Burak Turhan, Xiaofeng Wang, and Sira Vegas. 2021. Empirical Standards for Software Engineering Research. arXiv:2010.03525 [cs.SE]

[24] Ita Richardson and Yvonne Delaney. 2009. Problem based learning in the software engineering classroom. In *2009 22nd Conference on Software Engineering Education and Training*. IEEE, 174–181.

[25] Lucas Romao, Marcos Kalinowski, Clarissa Barbora, Allysson Alex Araújo, Simone D. J. Barbora, and Helio Lopes. 2024. Artifacts: Agile Minds, Innovative Solutions, and Industry–Academia Collaboration: Lean R&D Meets Problem-based Learning in Software Engineering Education. https://doi.org/10.5281/zenodo.10038021.

[26] Per Runeson, Martin Host, Austen Rainer, and Bjorn Regnell. 2012. *Case study research in software engineering: Guidelines and examples*. John Wiley & Sons.

[27] Yvonne Sedelmaier and Dieter Landes. 2015. Active and inductive learning in software engineering education. In *2015 IEEE/ACM 37th IEEE International Conference on Software Engineering*, Vol. 2. IEEE, 418–427.

[28] Williamson Alison Freitas Silva, Igor Fabio Steinmacher, and Tayana Uchôa Conte. 2017. Is It Better to Learn from Problems or Erroneous Examples?. In *2017 IEEE 30th Conference on Software Engineering Education and Training (CSEET)*. 222–231. https://doi.org/10.1109/CSEET.2017.42

[29] Bianca Rodrigues Teixeira, Bruna Ferreira, André Luiz Brandão de Damasceno, Simone DJ Barbosa, Cassia Novello, Hugo Villamizar, Marcos Kalinowski, Thuener Silva, Jacques Chueke, Hélio Lopes, et al. 2021. Lessons Learned from a Lean R&D Project.. In *ICEIS (2)*. 345–352.

[30] Mark Turner, Barbara Kitchenham, Pearl Brereton, Stuart Charters, and David Budgen. 2010. Does the technology acceptance model predict actual use? A systematic literature review. *Information and software technology* 52, 5 (2010), 463–479.

[31] Lev Semenovich Vygotsky and Michael Cole. 1978. *Mind in society: Development of higher psychological processes*. Harvard university press.

[32] Claes Wohlin, Aybuke Aurum, Lefteris Angelis, Laura Phillips, Yvonne Dittrich, Tony Gorschek, Hakan Grahn, Kennet Henningsson, Simon Kagstrom, Graham Low, Per Rovegard, Piotr Tomaszewski, Christine van Toorn, and Jeff Winter. 2012. The Success Factors Powering Industry-Academia Collaboration. *IEEE Software* 29, 2 (2012), 67–73. https://doi.org/10.1109/MS.2011.92